\documentclass[fleqn,10pt]{MySelfArx} 

\usepackage[english]{babel} 
\usepackage{csquotes}

\definecolor{color1}{RGB}{0,0,90} 
\definecolor{color2}{RGB}{0,20,20} 

\usepackage{hyperref} 

\hypersetup{
	hidelinks,
	colorlinks,
	breaklinks=true,
	urlcolor=color2,
	citecolor=color1,
	linkcolor=color1,
	bookmarksopen=false,
	pdftitle={Title},
	pdfauthor={Author},
}

\newcommand{\unit}[1]{\ensuremath{\mathrm{\,#1}}}
\renewcommand{\u}[1]{\unit{#1}}

\usepackage{amsmath}
\usepackage{hepnames}

\usepackage[utf8]{inputenc}
\usepackage[T1]{fontenc}  

\usepackage{upgreek}
\DeclareUnicodeCharacter{03BC}{\upmu}
\DeclareUnicodeCharacter{00B2}{\ensuremath{^2}}
\DeclareUnicodeCharacter{00B3}{\ensuremath{^3}}
\DeclareUnicodeCharacter{03A9}{\ensuremath{\Omega}}
\DeclareUnicodeCharacter{00D7}{\ensuremath{\times}}

\JournalInfo{Submitted to NIM A} 
\Archive{Contribution to PISA'22 proceedings} 

\PaperTitle{New results from the technological prototype \\
	of the CALICE highly-granular silicon tungsten electromagnetic calorimeter} 

\Authors{V.~Boudry for the CALICE collaboration\textsuperscript{1}*} 
\affiliation{\textsuperscript{1}\textit{Laboratoire Leprince-Ringuet, CNRS, École polytechnique, Institut Polytechnique de Paris, 91120 Palaiseau, France.}} 
\affiliation{*\textbf{Corresponding author}: vincent.boudry@in2p3.fr} 

\newcommand{\keywordname}{Keywords} 

\Abstract{An extended version of the CALICE silicon-tungsten ECAL, with 15 active layers, was tested in November 2021 at the DESY beam test facility. %
  Respectively with the previous design, it featured some complete layers with a thin PCB design, and a new compact DAQ handling 15 layers in a common set. %
  The noise, self-trigger performance and response to punch-though electrons, without tungsten absorbers, was evaluated for each of the 15,360 channels,
  using an improved procedure.
  Showers of low-energy electrons (1 to 5\u{\mathsf{GeV}}) have been recorded in two configuration of absorber, and are being analysed. %
  Some preliminary results will be shown here.
}

\begin{document}

\maketitle 

\tableofcontents 

\thispagestyle{empty} 



\section{Introduction} 
\subsection{Particle Flow Detectors at Higgs Factories}

Detectors for future Higgs factories, near any $\mathrm{e^+ e^-}$ collider (ILC, FCC-ee, CEPC or CLIC), will require an excellent energy resolution for jets ($\sim 3\text{--}4\u{\%}$ for WW or ZZ di-bosons masses), 
particle identification, large acceptance and a sensitivity to unconventional physics, such as long-lived particles. %
A major path toward these goals is the use of highly-granular calorimeters, which, combined with a precision tracker, allows the optimal use of Particle Flow Algorithms (PFA)~\cite{thomson_particle_2009}. %

Particle identification, which improves the PFA performance, usually provided by the tracker $dE/dx$ for low momentum particles, can be enriched by the analysis of shower shapes, and by the measurement of the time-of-flight in the ECAL. %
With the studies for the High-Luminosity LHC -- and more specifically the CMS HGCAL endcap upgrade program -- the large potential of precision timing (few 10\u{ps}) for the particle identification and event filtering has been revealed. %
This performance is linked to the small cell size ($30\u{ps} \sim 1\u{cm}/c$) in those devices.

\subsection{An Ultra-Granular SiW-ECAL for Higgs Factories}

The requirements of imaging capacities for an ECAL add to the standard calorimetric ones the handling of a very large number of channels, and the need to have compact devices (to stem narrow showers) to better separate nearby particles. %
The ECAL must also have a large dynamic range at the cell level, to detect and track particles at their minimum of ionization (1~MIP) and measure electrons and photons from a fraction of GeV to several hundreds of GeV (a few 1000s MIPs in a cell).

Highly resistive Silicon sensors made of large area of segmented PIN diodes fulfil all the listed requirements needed (flexibility in the design, sensitivity, stability, speed), at a cost slightly larger than other solutions, but providing also an intrinsic stability of calibration. %
Associated with tungsten as a dense absorber, and a carbon fibre holding structure, it offers compactness and uniformity by minimizing dead spaces. %

For a SiW-ECAL using silicon as sensor and tungsten as absorber near the ILC, for which this R\&D was initiated, an optimal configuration was found~\cite{thomson_particle_2009} by having $5×5\u{mm²}$ cells and 30 layers in depth, for 24\u{X_0}. %
This reflects in having a channel density of $\sim 600\u{channel/dm³}$, and typically 100 millions channels in total for the ECAL alone. %
The readout electronics must then be integrated inside the detector, to multiplex the data output, and have a low power consumption. %
Their support boards, connecting them to the sensors, must be designed in the perspective of the production of 100,000's of elements.

The full design of the ILD SiW-ECAL is described in~\cite{behnke_international_2013-4} and~\cite{the_ild_collaboration_international_2020}.

\subsection{CALICE prototypes for ILC}

The CALICE collaboration aims at developing and testing imaging calorimeters for the Higgs Factories.  %

Following the building and testing of SiW-ECAL “physical prototype''~\cite{repond_design_2008}, a “technological prototype'' is in the making, which is the subject of this contribution.  %
It features chained embedded readout SKIROC2 ASICs, mounted on front-end boards, dubbed FEVs, onto which 4 silicon sensors are glued: this detection unit is called an ASU (for Assembly Single Unit). %
Compared with the physical prototype, the channel density has been increased by a factor 4, with cells sizes of $5×5\u{mm²}$, and the square silicon diode matrices enlarged to $9×9\u{cm²}$ (inscribed in 6'' circular wafers). %
The design of the FEV is complex, as it concentrate many constraints : holding of analogue and digital signals, the thinnest possible imprint, scalability and mechanical precisions. %
A number of layers have been produced, with different designs, aiming at a final number of 26 to 30, but not yet fulfilled.
The production of a working $18×18\u{cm²}$ calorimeter is a key milestone, nearly achieved. %

A complementary milestone, not detailed here, will be the operation of 8 ASUs daisy-chained into a long cassette, as the base element for a large SiW-ECAL.

\paragraph{The SKIROC2 ASIC:}
the SKIROC2 ASIC, designed by the Omega lab~\cite{suehara_performance_2018} for an SiW-ECAL near the ILC, pioneered the embedded readout of highly-granular silicon calorimeters\footnote{~it was used for the early prototype of HGCAL, then derived as SKIROC-CMS, and finally fully redesigned as HGCROC.}.  %
The signal of each of 64 channels is first pre-amplified and send to 3 branches: 
\begin{itemize}
	\item one with a fast shaping ($\sim 20\u{ns}$) for triggering and timing purposes, and 
	\item two with a slow shaper ($\sim 200\u{ns}$) for amplitude measurements, using gains of 1 and 10.
\end{itemize}
On the crossing of an adjustable threshold for any of the 64 channels, the amplitude of the $64×2$ slow shapers or optionally\footnote{~2 out of 3 values can be selected} of a slow shaper plus a TDC ramp will be stored in an analogue memory. The low or high gain channels can be selected according to a programmable threshold. %
At the end of an acquisition period, corresponding to a collider spill, the $2×15$ analogue memories per channels, are converted using 12-bit Wilkinson ADCs, and tagged with a 
`Bunch-crossing' ID (BCID), counting the periods spent between the spill start and the trigger.

To minimize its power consumption, the SKIROC2 can be power-pulsed by parts, to be operated only during the acquisition and conversion phases, amounting to about 1\u{\%} of the total at the ILC.

A second version of the ASIC, the SKIROC2a, was produced, improving on noise, clock handling and threshold adjustments.

\paragraph{The integration challenge:}
most of the technological constraints concentrate on the FEV design, as the main piece of the ASUs: to fulfil the scaling, compactness and precision, the ASU is the base unit on which the detector will be built. %
Between 8 and 12 ASUs will be assembled into cassettes, of 1.5 to 2.2\u{m} long, operated from a single end to ensure the detector uniformity. %
The FEV ensures power distribution, clock, and configuration dispatching to the ASICs, and their data collection, using daisy-chaining to minimize the number of transmissions lines (140 at this date). %
This must be done without compromising the weak signal (2\u{fC} to 10\u{pC}) transmission from the sensors to the ASIC. %
Additionally, the FEV thickness must be kept low, of the order of 2\u{mm} ASIC and ancillary electronics included, in order to keep the detector compact.
Finally, a mechanical precision of the order of 50\u{μm} is needed to keep the ASUs properly aligned, and to cover the fragile sensors without providing gaps between the ASUs. %
The flatness of the PCBs is also checked, to avoid breaking the sensors glued to them.
As for much of the electronics, the connectors are a critical component, for which specific R\&D was required. 

The FEVs currently used in the prototype come in two flavours: %
\begin{itemize}
\item{BGA}: one in which the ASICs are packaged, using Ball Grid Array (BGA) connections. %
This allows for pre-mounting testing of the ASICs and an easier handling, at the expense of the thickness: $\sim1.2\u{mm}$ are needed, which is also used for connectors, filters, etc.
Versions 10 to 13 of these boards are in use, with 2 versions of the ASICs and 3 thicknesses of sensors, as described in table~\ref{tab:FEV}.
FEV10 and 11 are almost identical, and are treated as a single type. 
The FEV12s differ only lightly in the PCB design, but benefits from SKIROC2a's. 
The FEV13 is a major redesign of the PCB, especially for the routing of sensor signals to the ASICs. 
\item{COB:} The original design embeds the naked ASICs inside the PCB, with a direct bonding connection, are dubbed COB (for Chip-on-Board). 
The additional difficulties (mechanical constraint, absence of decoupling) have just been lifted.
The first two equipped working versions of the COBs, based on FEV11, have been employed. 
\end{itemize}

\begin{table}[h]
\begin{center}
	\begin{tabular}{|l|c|c|c|}
	\hline
	PCB       &   ASIC   &  Sensor   & Number \\ \hline
	FEV10, 11 & SKIROC2  & 320\u{μm} &  1+6   \\ \hline
	FEV12     & SKIROC2a & 500\u{μm} &   4    \\ \hline
	FEV13     & SKIROC2a & 650\u{μm} &   2    \\ \hline
	FEV11-COB & SKIROC2a & 500\u{μm} &   2    \\ \hline
	\end{tabular}
\end{center}
	\caption{Summary of the ASUs compositions}
	\label{tab:FEV}	
\end{table}

\paragraph{A compact DAQ:}
a new DAQ system\cite{breton_calice_2020-1} has lately been set up, connecting 15 layers on a single readout Kapton, through low occupancy interface boards (SLBoard). %
For the final detector, a minimal space, of about 3\u{cm}, is foreseen between the Hadronic Calorimeter and the ECAL. %
This is demanded by the particle tracking between the two devices. %
All services (power, cooling, DAQ, mechanical support) must take place in this space. %
At the end of cassettes, a service board, dubbed SLboard, ensures the low- and high-voltage powering, DAQ clock and control distribution.
It presents a longitudinal span of less than 6\u{cm}, close to the design of 4.3\u{cm} ($3\u{cm} / \cos(45°)$), a height compatible with a layer spacing of 15\u{mm}, and leaves space for the active cooling pipe at the centre.
\begin{figure}
	\includegraphics[width=\linewidth]{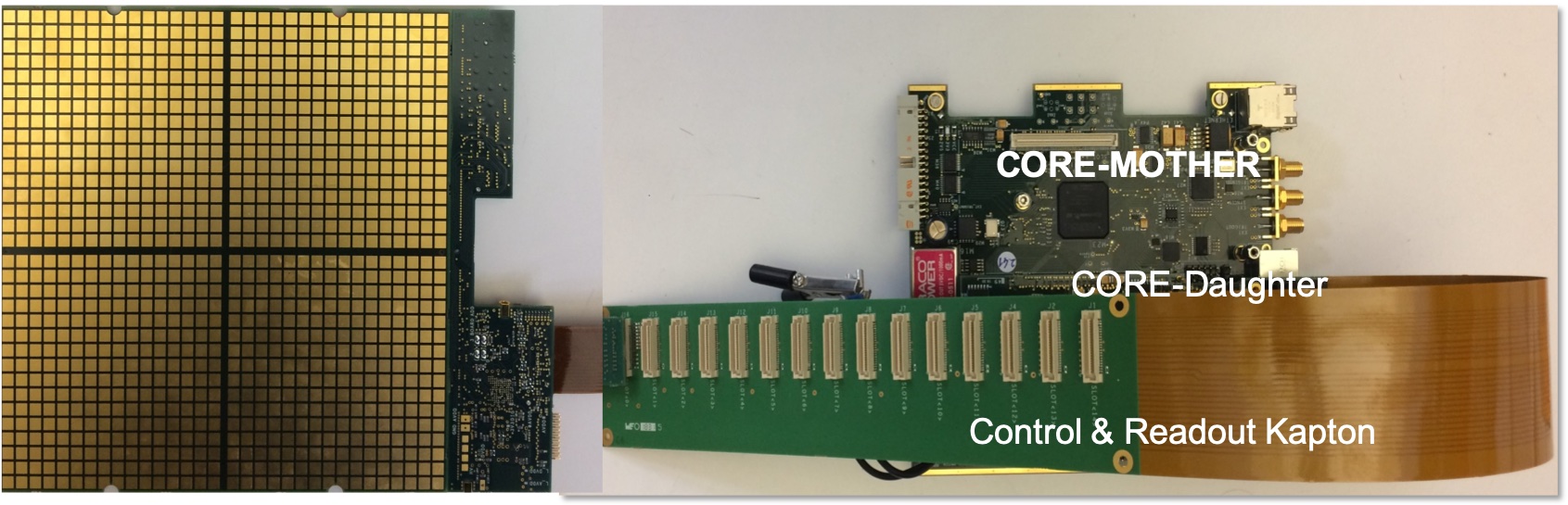}
	\caption{Picture of the complete DAQ attached to a single layer.}
\end{figure}

Up to 15 SLboards can be operated by a Kapton connected to a mezzanine board on a `CORE' module, which has place for two of them.

A labWindows based acquisition and monitoring application controls and read the CORE module via a single USB2 cable. %
The data can be partially analysed online to provide basic or advanced distributions (number of hits per layer, hit maps, amplitude, and time distribution… ) for optimal or reduced acquisition speeds.
An interface to the EUDAQ acquisition software developed for AIDA \& AIDA-2020~\cite{ahlburg_eudaq-data_2020}, is provided; it allows for combined running with other compatible devices, such as the EUDET/AIDA telescope, or the CALICE AHCAL.

\section{Data taking at the DESY beam test facility}
\subsection{Beam characteristics}

The DESY beam test facility~\cite{diener_desy_2019} offers beams of electrons or positions, between 1 and 6\u{GeV}, with a peak intensity of a few kHz around 3\u{GeV}.
The beam spot on the detector, installed in the T21 zone on a movable table, had a Gaussian profile of about 10\u{mm}.

Four weeks of test have been used in November 2021 and March 2022.  %
About three weeks were dedicated to the commissioning and training of the system, adjusting detector electronics parameters, calibration and DAQ parameters and monitoring tools. 

\subsection{The prototype layouts}

\begin{figure}
	\begin{tabular}{cc}
	\end{tabular}
	\begin{center}
		\includegraphics[width=0.90\linewidth]{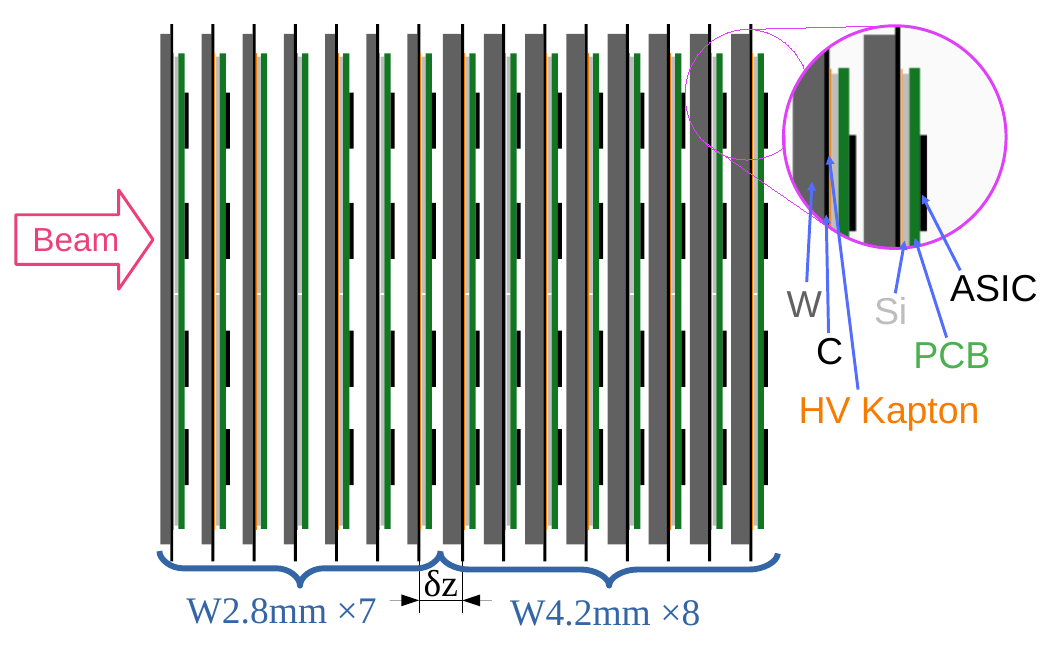}
	\end{center}	
	\caption{ 
		A detection layer consists of one ASU connected to a SLboard. It is mounted on a 2\u{mm} Carbon Fibre Reinforced Polymer plate, equipped with 3D printed rails supports.
		The associated tungsten plate can be mounted on the other side of the carbon plate. 
		A schematic of the layers dispositions for the DESY March 2022 beam set-up.  The layers are equally spaced every $\delta z = 15\u{mm}$.}
	\label{pic:stack}
\end{figure}

During this campaign, for the first time 15 detection layers were operated together, in a calorimeter configuration allowing for partial or complete containment of the electromagnetic showers (see Fig.~\ref{pic:stack}).
The detection layers are mounted in a protective case, providing a fix spacing of 15\u{mm}, lateral alignment, and light and CEM protection.

\begin{figure}
	\begin{tabular}{c}
	\includegraphics[width=0.95\linewidth]{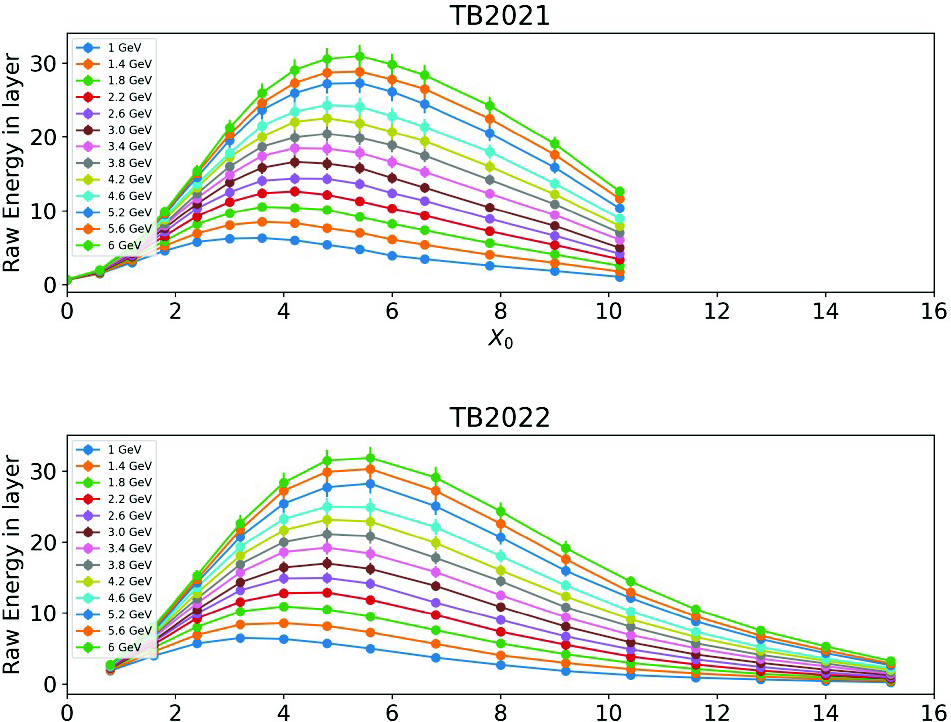}
	\end{tabular}
\caption{Simulated electromagnetic shower profiles, scaled in units of \u{X_0} expected for
		the Nov21 (top) and Mar22 (bottom) tungsten configurations.}
	\label{pic:EMProfile}		
\end{figure}

Three tungsten absorber configurations have been employed, with (seen from the beam entrance):
\begin{itemize}[font=\bfseries,leftmargin=1.3cm]
	\item[noW:] no tungsten;
	\item[Nov21:] 11 layers of 2.1\u{mm} (0.6\u{X_0}) plus 3 layers of 4.2\u{mm} (1.2\u{X_0}) for 10.2\u{X_0} in total;
	\item[Mar22:] 7 layers of 2.8\u{mm} (0.8\u{X_0}) plus 8 layers of 4.2\u{mm} (1.2\u{X_0}) for 15.2\u{X_0} in total;
\end{itemize}
The first configuration ("noW”) let the electrons passing almost unhindered, providing a pseudo minimum-ionizating particle (MIP) signal.
The second configuration was used in November 2021, the third in March 2022.
The expected simulated shower profiles in units of \u{X_0} are shown in Fig.~\ref{pic:EMProfile}.

\section{Noise studies}

The low occupancy of highly-granular calorimeters operated at lepton colliders allows for local triggering.  %
For pulsed linear colliders (not so for continuous circular ones), the data can be locally stored until the end of a bunch spill, converted and readout outside the acquisition period. %
This operating mode present advantages to reduce the power consumption, but renders the electronics highly sensitive to noisy channels (and collective modes), which by filling the local memories effectively blind their ASICs neighbours. %
The tuning of the threshold (common to the 64 channel in the SKIROC2) and masking of the noisier channels is mandatory, and can only be done with beam data~\footnote{~the signal-to-noise ratio is vastly different in cosmic and beam operations, especially due to pulsing operation.}.

\begin{figure}[t!]
	\begin{tabular}{c}
		\includegraphics[width=0.8\linewidth]{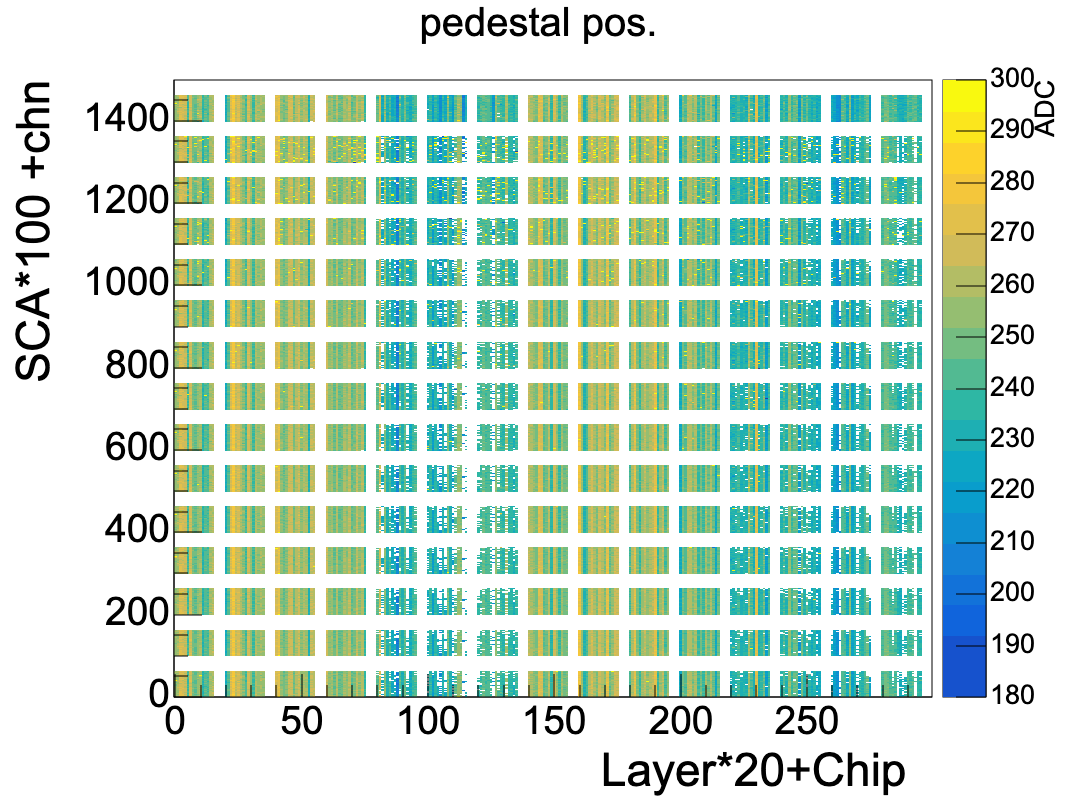}\\
		\includegraphics[width=0.83\linewidth]{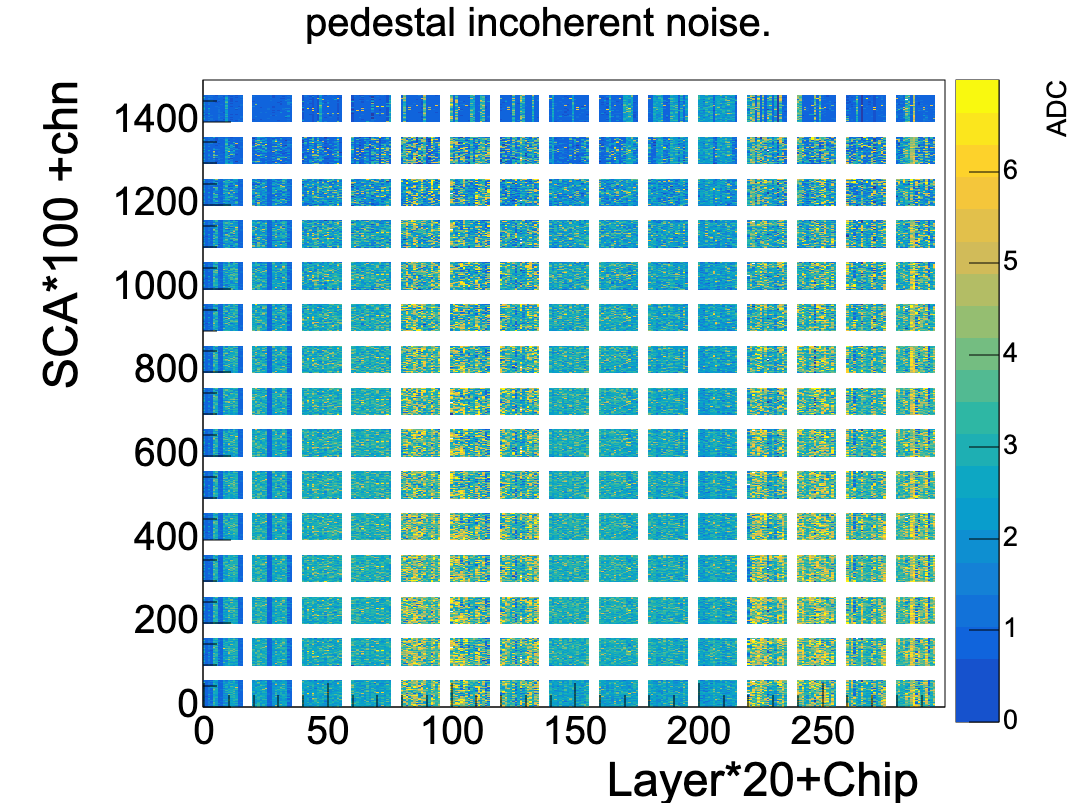}
		\\
	\end{tabular}
	\caption{ Maps of the pedestal (top) and incoherent noise (center) in units of ADC counts for each of the 230,400 memory. The horizontal axis is a unique ID for the ASICs for all layers. The vertical axis is unique ID for the channels and analogue memories (SCA)
	}
	\label{pic:noise-maps}
\end{figure}

The partial zero-suppression is made by the SKIROC2 ASICs: any channel above threshold, defined as a `hit', triggers the readout of the 64 channels. This ensures the recording of many noise signal in both high and low-gain branches.
These were analysed in terms of coherent and incoherent sources, using the framework of~\cite{frisson_coherent_2014}: assuming Gaussian noise sources, the inter-channel correlation allows quantifying the different contributions.
The 230,400 memories (15 layers of 1024 channels, each with 15 memory slots) have been analysed, the pedestal and incoherent noise extracted (see Fig.~\ref*{pic:noise-maps} for the high-gain branch).
The pedestal distribution exhibits primarily a per-chip correlation, due to the ASICs technological analogue dispersion. %

\begin{figure}[t]
	\hspace{-0.4cm}
	\begin{tabular}{c}
		\includegraphics[width=0.360\linewidth]{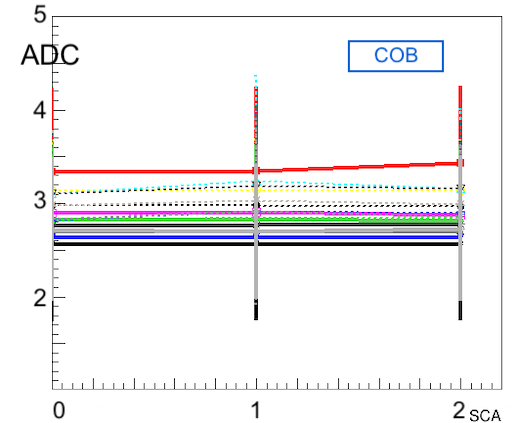}
		\includegraphics[width=0.320\linewidth]{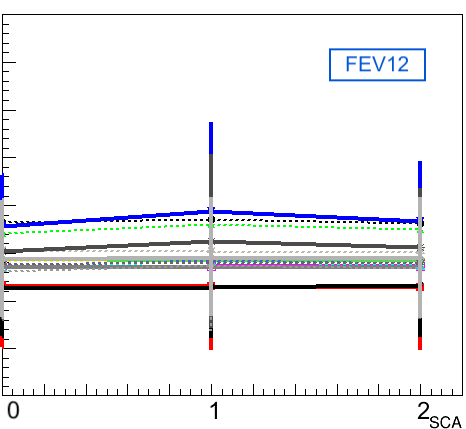}
		\includegraphics[width=0.320\linewidth]{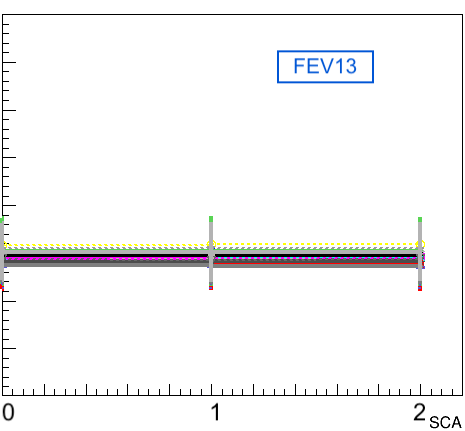}
	\end{tabular}
	\caption{
		Average and dispersion of incoherent noise measurements on the 64 channels, for the 3 first memories (SCA), for each of the 16 ASIC of 3 types of boards: COB (left), FEV12 (middle), FEV13 (right).\\
		Compared to the FEV12, the COB suffers slightly from the near lack of decoupling capacitors. The optimization of signal routing on the FEV13, based on observations on the previous versions, is clearly visible.}
	\label{pic:noise}
\end{figure}

The incoherent noise is found in the 2.5 to 3.5 ADC counts range, depending essentially on the FEV type (see Fig.~\ref*{pic:noise}). This must be compared to the signal of a MIP, ranging between 70 and 140 ADC counts for 320 and 650\u{μm} sensors respectively~\footnote{~this translates to SNR of 20 to 40 for that branch.}, and is similar to previous results using individual fitting of the channels~\cite{kawagoe_beam_2019}.

The noise pattern is now well identified, with a clear influence of power lines near the SLboards, a recurring problematic channel number (37) on all ASICs, occasionally affecting its neighbours, plus some random rogue ones.
The FEV13 design tackled most of the issues (but the channel 37).

\section{Response to MIPs}

\begin{figure}
	\centering
	\includegraphics[width=0.7\linewidth]{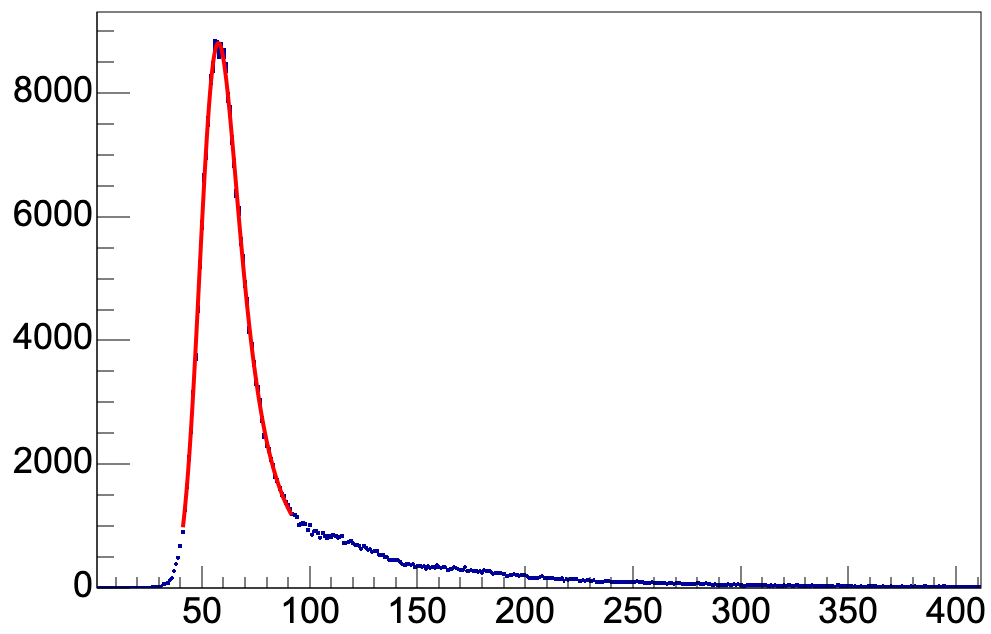} 
	\includegraphics[width=0.7\linewidth]{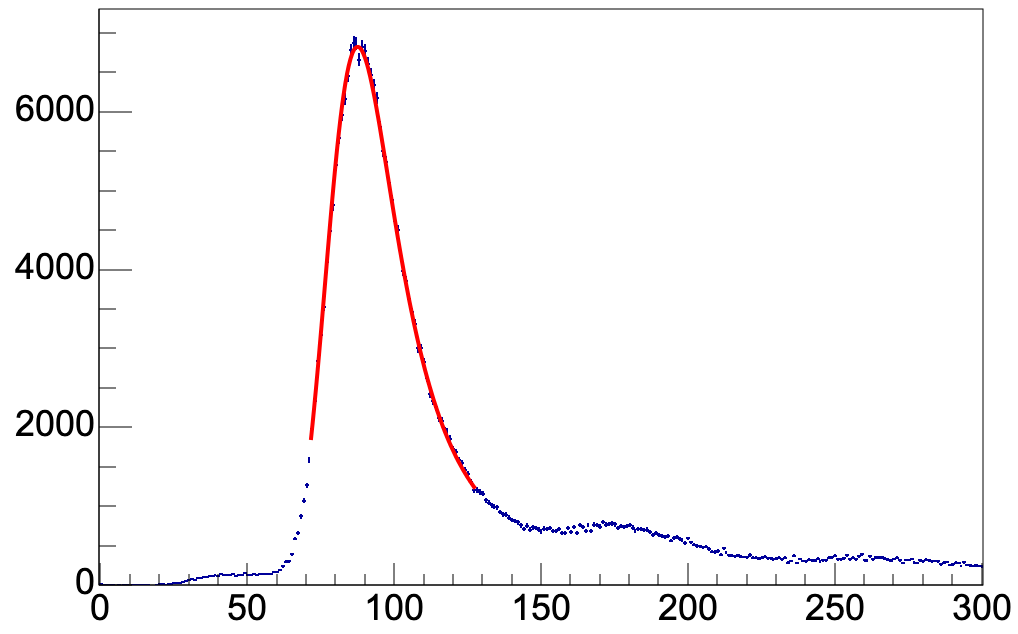}
	\caption{Examples of energy spectra recorded in the cells of a COB layer (top) at the front and of a FEV12 (bottom) in the back of the detector, without tungsten absorber. The horizontal axis are in units of ADC counts, pedestal subtracted. The vertical axis are events count per bin. The higher fraction of 2 and 3 MIPs, compared to the predominant single MIP peak, in the back stem from interactions in the foremost layers. The fitting (red lines) of the single MIP peak is used to calibrate the detector cells.}
	\label{pic:mips}		
\end{figure}

The inter-calibration of the prototype is performed by a positional scan with all the tungsten removed.
The observed spectra in every cell, can be nicely be fitted by a Landau distribution convoluted with a Gaussian (see Fig.~\ref*{pic:mips}) in most cases.
Simulation studies have shown no obvious difference in MIP spectrum stemming from a 3\u{GeV} electron and a muon, but for the presence of a small two-MIP peak in the later layers for the electrons.

The preliminary overall picture shows a  response in many layers uniform at better than 30\u{\%}, dominated by the fitting procedure. The response is strongly correlated with the thickness of sensors; the relative response for thick sensors (500 and 650\u{μm}) is about $0.18\pm0.05\u{adc/μm}$, it is slight larger for 320\u{μm} sensor at $0.21\pm0.05\u{adc/μm}$.  
But some layers feature a pattern of missing signal (statistics too low to perform fits) at some edge of some sensors.  %
The pattern suggest a weakness of the conductive glue dots connecting the sensors to the FEVs readout pads, which is under investigation.

\begin{samepage}
\section{Response to low-energy electrons}
\subsection{Simulation}
The geometry is described using the DD4HEP framework, providing a flexible geometry overlay on GEANT4.
\end{samepage}
The quantum of energy deposition provided by GEANT4 in each of the cell of the model will be processed in a “digitization'' module which mimics the treatment in the SKIROC2 ASIC.
The fast and slow branches shaping are replicated, a threshold applied on the former, providing a trigger time, which added by a fixed delay is used to sample the latter. %
The data is calibrated on MIPs as for the real data.  %
The fine parameters (shaping form, noise) of the module are being adjusted on the data.

The cells masked during the acquisition are also masked in the simulation.

\subsection{In-shower cell energy spectrum}

\begin{figure*}[t!]
	\centering
	\includegraphics[width=0.9\linewidth]{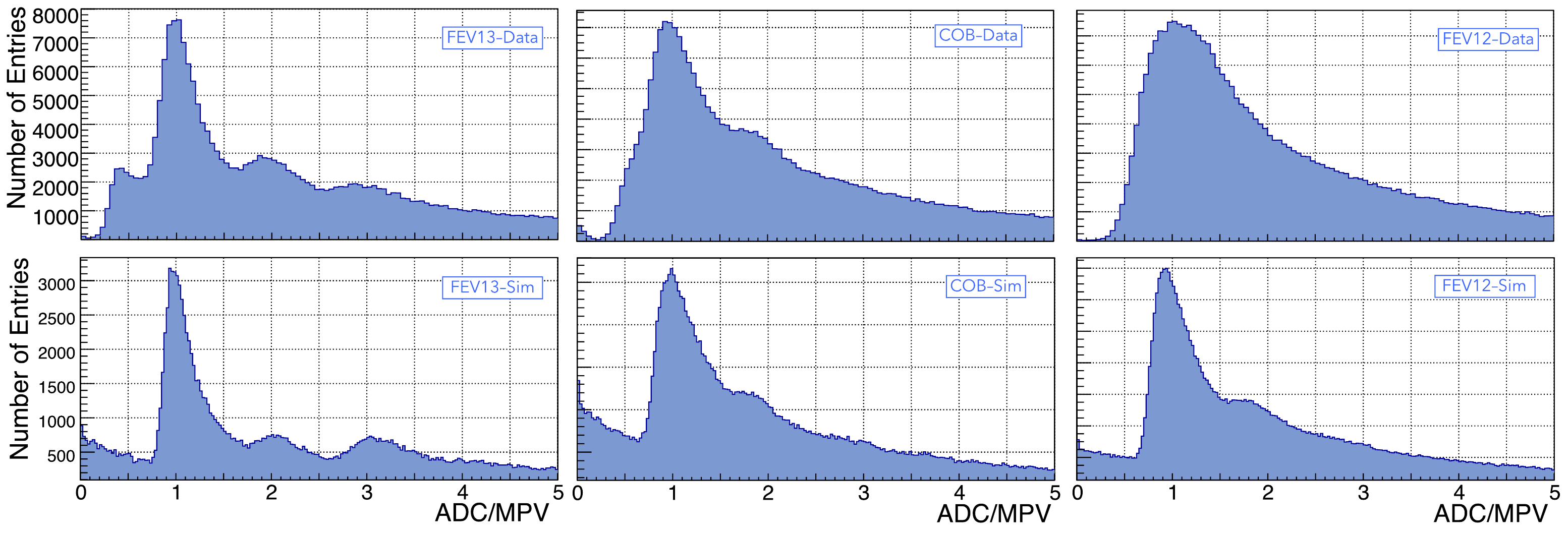}
	\caption{
		Energies deposited in cells of an FEV13 equipped with 650\u{μm} sensors (left), a COB with 500\u{μm} sensors (middle) and a FEV11 with 320\u{μm} sensors,
		scaled to the Most Probable Value for a single particle crossing the cell. %
		On top the spectra from data, at the bottom simulated ones, without digitization or built-in noise. %
		The low-energy tails stem from charge splitting between two cells 
    }
	\label{pic:in-shower}		
\end{figure*}

The energy deposited in 3 types of boards at 3 depths in 3\u{GeV} electron showers is displayed in Fig.~\ref*{pic:in-shower}. %

Despite the difference of positions and quality of the boards, by comparing the response, the following conclusion can be drawn.
The contributions from singles, pairs, or triplets of particles can be clearly separated in the FEV13-based ASU, placed at the front of the calorimeter, after 1.6\u{X_0}.
This is well reproduced by the raw simulation (no digitization applied). 
It can be observed that the noise contribution, which should enlarge the peaks, is small. %
The effective threshold can unambiguously be determined, in this case at a value of about 1/4th of the most probable value (MPV) of a MIP~\footnote{~for a 320\u{μm} sensor the MPV is about 70 ADC counts above the pedestal}, with a small dispersion.

At the place of the COB based ASU, placed at 3.2\u{X_0} from the front, only the single particle and pairs can be seen.  %
A threshold can be estimated around 0.4\u{MPV}.

Past the maximum of the shower (6.8\u{X_0}), the single- and two-particle peaks are still visible in the simulated data, but are washed out by the noise on the FEV11-based ASU.  %
In this case, even the identification of the main peak in the data is unclear: it could be a single-particle with washed-out 2-particle peak, or the 2-particle peak, with the single-particle one cut-out by the electronics threshold. %
Only the comparison with neighbouring layers and scaling allows deciding for the first hypothesis.  %
An effective threshold around 2/3\u{MPV} can be estimated, cutting part of the foot of first peak.

Provided a sufficient signal-to-noise ratio, in-shower calibration of the response to a MIP and position of the threshold can thus be performed.

A saturation of the cell energy in the high-gain branch was observed for the ASUs equipped with 500\u{μm} or 650\u{μm} sensors for 3\u{GeV} electrons. %
The margin was large enough to ensure linearity in the low-gain branch even for 6\u{GeV} particles, but the gain must be adjusted for higher beam energies, e.g\@. at the CERN facility.
Configuring a higher feedback capacitor in the SKIROC2 pre-amplication stage (6\u{pF} instead of 1.2\u{pF}) lowered the overall gain of the ASICs by 4.73, as extracted from MIP calibration.

\section{Outlook}

The data sets recorded at the DESY beam test facility in November 2021 and March 2022, with the first version of the SiW-ECAL prototype featuring enough layers to perform a quantitative analysis of its calorimetric response, are being analysed. %
Some preliminary results on calibration and per-cell behaviours have been shown, not yet including estimations of the energy resolution, as this requires full calibration and the proper handling of the detectors features (e.g. masked channels).
Noise and punch-through electrons have yielded calibrations for the equalization of the response of the detector and masking of defective channels.
The detector is mostly reacting as expected, with per-layer calibration sets compatible with the previous campaigns, but some layers show weaknesses in the gluing dots near the edge of some sensors.

The electrons shower per-cell energy deposition is roughly reproduced by the simulation; this will be refined by tuning and applying the digitization mimicking the SKIROC2 signal processing.
For layers with a proper signal-to-noise ratio, energy calibration and threshold value of individual cells can be extracted directly from the shower data.

The highest possible electronics gain, used so far to study the noise and low-energy response, saturates some cells in 3\u{GeV} electron showers.
For the next campaign at CERN in June 2022, this gain will be reduced, some ageing FEV11-based ASUs will be replaced by FEV13-based ones, and the amount of tungsten absorber thickness will be increased to 18.4\u{X_0} to better contain high-energy electron showers.

Finally, the behaviour of existing layers has provided input to correct the design of the next generation of boards, to be produced by the end of 2022, improving on power and signal distribution, stabilization, flexibility, and HV filtering.

\section{Acknowledgements}

This project has received funding from the European Union’s Horizon 2020 research and innovation programme under grant agreement No 101004761.
The measurements leading to these results have been performed at the Test Beam Facility at DESY Hamburg (Germany), a member of the Helmholtz Association (HGF).
We are very thankful to the DESY beam test facility operation teams~\cite{diener_desy_2019}.


\phantomsection
\bibliographystyle{utcapsnoxref}
\bibliography{CALICE_siwecal}


\end{document}